\overfullrule 0pt
\input epsf.tex
\def\bigskip{\vskip 3mm}

\nopagenumbers

\font\rfont=cmr10 at 10 true pt
\def\ref#1{{\hbox{{ [#1]}}}}


\font\fourteenbf=cmbx12 scaled\magstep1

\font\tenbfit=cmbxti10
\font\sevenbfit=cmbxti10 at 7pt
\font\fivebfit=cmbxti10 at 5pt
\newfam\bfitfam \def\bfit{\fam\bfitfam\tenbfit}
\textfont\bfitfam=\tenbfit  \scriptfont\bfitfam=\sevenbfit
\scriptscriptfont\bfitfam=\fivebfit

\def\a {\alpha}   
\def\e{\epsilon}  \def\o{\omega}
\def\s {\sigma}

\def\pmb#1{\setbox0=\hbox{#1}
 \kern.05em\copy0\kern-\wd0 \kern-.025em\raise.0433em\box0 }

\def\slash{/\kern-.5em}


 %


\def\boxit#1{\vbox{\hrule\hbox{\vrule\kern1pt\vbox
{\kern1pt#1\kern1pt}\kern1pt\vrule}\hrule}}

\def\h{\hfill\break}
\parskip=6pt
\parindent=0pt
\hsize=12truecm
\vsize=19truecm
\def\footnoterule{\kern-3pt
\hrule width 17truecm \kern 2.6pt}


\catcode`\@=11 

\def\nolabels{\def\wrlabeL##1{}\def\eqlabeL##1{}\def\reflabeL##1{}}
\def\writelabels{\def\wrlabeL##1{\leavevmode\vadjust{\rlap{\smash%
{\line{{\escapechar=` \hfill\rlap{\sevenrm\hskip.03in\string##1}}}}}}}%
\def\eqlabeL##1{{\escapechar-1\rlap{\sevenrm\hskip.05in\string##1}}}%
\def\reflabeL##1{\noexpand\llap{\noexpand\sevenrm\string\string\string##1}}}
\nolabels
\global\newcount\refno \global\refno=1
\newwrite\rfile
\def\defref{{{\hbox{ [\the\refno]}}}\nref}
\def\nref#1{\xdef#1{\the\refno}\writedef{#1\leftbracket#1}%
\ifnum\refno=1\immediate\openout\rfile=refs.tmp\fi
\global\advance\refno by1\chardef\wfile=\rfile\immediate
\write\rfile{\noexpand\item{#1\ }\reflabeL{#1\hskip.31in}\pctsign}\findarg}
\def\findarg#1#{\begingroup\obeylines\newlinechar=`\^^M\pass@rg}
{\obeylines\gdef\pass@rg#1{\writ@line\relax #1^^M\hbox{}^^M}%
\gdef\writ@line#1^^M{\expandafter\toks0\expandafter{\striprel@x #1}%
\edef\next{\the\toks0}\ifx\next\em@rk\let\next=\endgroup\else\ifx\next\empty%
\else\immediate\write\wfile{\the\toks0}\fi\let\next=\writ@line\fi\next\relax}}
\def\striprel@x#1{} \def\em@rk{\hbox{}} 
\def\lref{\begingroup\obeylines\lr@f}
\def\lr@f#1#2{\gdef#1{\defref#1{#2}}\endgroup\unskip}
\newwrite\lfile
{\escapechar-1\xdef\pctsign{\string\%}\xdef\leftbracket{\string\{}
\xdef\rightbracket{\string\}}}

\def\writestop{\def\writestoppt{\immediate\write\lfile{\string\p
ageno%
\the\pageno\string\startrefs\leftbracket\the\refno\rightbracket%
\string\def\string\secsym\leftbracket\secsym\rightbracket%
\string\secno\the\secno\string\meqno\the\meqno}\immediate\closeout\lfile}}
\def\writestoppt{}\def\writedef#1{}
\catcode`\@=12 

\centerline{\fourteenbf SOFT INTERACTIONS}

\vskip 5mm
\centerline{A Donnachie}
\vskip 2mm
\centerline{Department of Physics and Astronomy, University of Manchester}
\vskip 4mm
\centerline{P V Landshoff}
\vskip 2mm
\centerline{DAMTP, University of Cambridge}
\vskip 5mm

Existing data on total cross sections, on elastic scattering at small, moderate
and large values of $t$, and on diffraction dissociation, reveal a surprisingly
simple phenomenology, but they throw up many
questions for the LHC to answer.
\bigskip
{\bf Total cross-sections}

It is common ground that soft interactions are described by Regge theory%
\defref\collins{
P D B Collins, {\it Introduction to Regge Theory},
Cambridge (1977)
}, but the details are still controversial. Our own attitude%
\defref\total{
A Donnachie and P V Landshoff, 
Physics Letters B296 (1992) 227}
is to adopt the simplest approach as far as the data allow. 
In its crudest form, this approach results in a fit to total cross-section
data with just two fixed powers, as is shown in figure 1.
While most fits to the existing data predict a cross-section at LHC energy
not far from 110 mb, and this agrees with a cosmic-ray measurement%
\defref\pdg{
Particle Data Group, Physical Review D54 (1996) page 193
}, the conflict between the two Tevatron data points%
\defref\tev{
E710 collaboration: N Amos et al, Phys Lett B243 (1990) 158\h
CDF collaboration: F Abe et al, Physical Review D50 (1994) 5550
}
in figure 1 raises the question whether it might be considerably larger.
An explanation for such a larger values, if it were to be found, would be
readily available: the BFKL pomeron\defref\bfkl{
E A Kuraev, L N Lipatov and V Fadin, Soviet Physics JETP 45 (1977) 199\h
J R Cudell, A Donnachie and P V Landshoff, Nuclear Physics BB482 (1996) 241
}.
\bigskip
{\bf Elastic scattering -- small} {\bfit t}

In the fits shown in figure 1, the term $s^{\e}$, with $\e \approx 0.08$, 
is said to be associated with the exchange of the soft pomeron. Although
the fits work well with constant $\e$, it cannot really be constant.
This is because one must take account not only of the exchange of a single 
pomeron, but also of more than one. It may be that single-pomeron exchange
yields a fixed power $s^{\e _0}$ (though even this is not certain),
but including the the exchange of two pomerons reduces  the effective
power and makes it vary with $s$. The fact that the data fit well to constant
$\e$ suggests that its variation with $s$ is very slow, so that the
contribution to the total cross-section  from the exchange of two
or more pomerons is rather small and $\e _0$ is only a little greater
than 0.08. This is the approach that we have taken, though others%
\defref\capella{
A Capella et al, Physics Letters B337 (1994) 358\h
E Gotsman, E M Levin and U Maor, Physical Review D49 (1994) 4321
} disagree: for them, $\e _0$ is significantly greater than 0.08 and
multiple exchanges are not small.
\topinsert
\centerline{\epsfxsize 100truemm\epsfbox{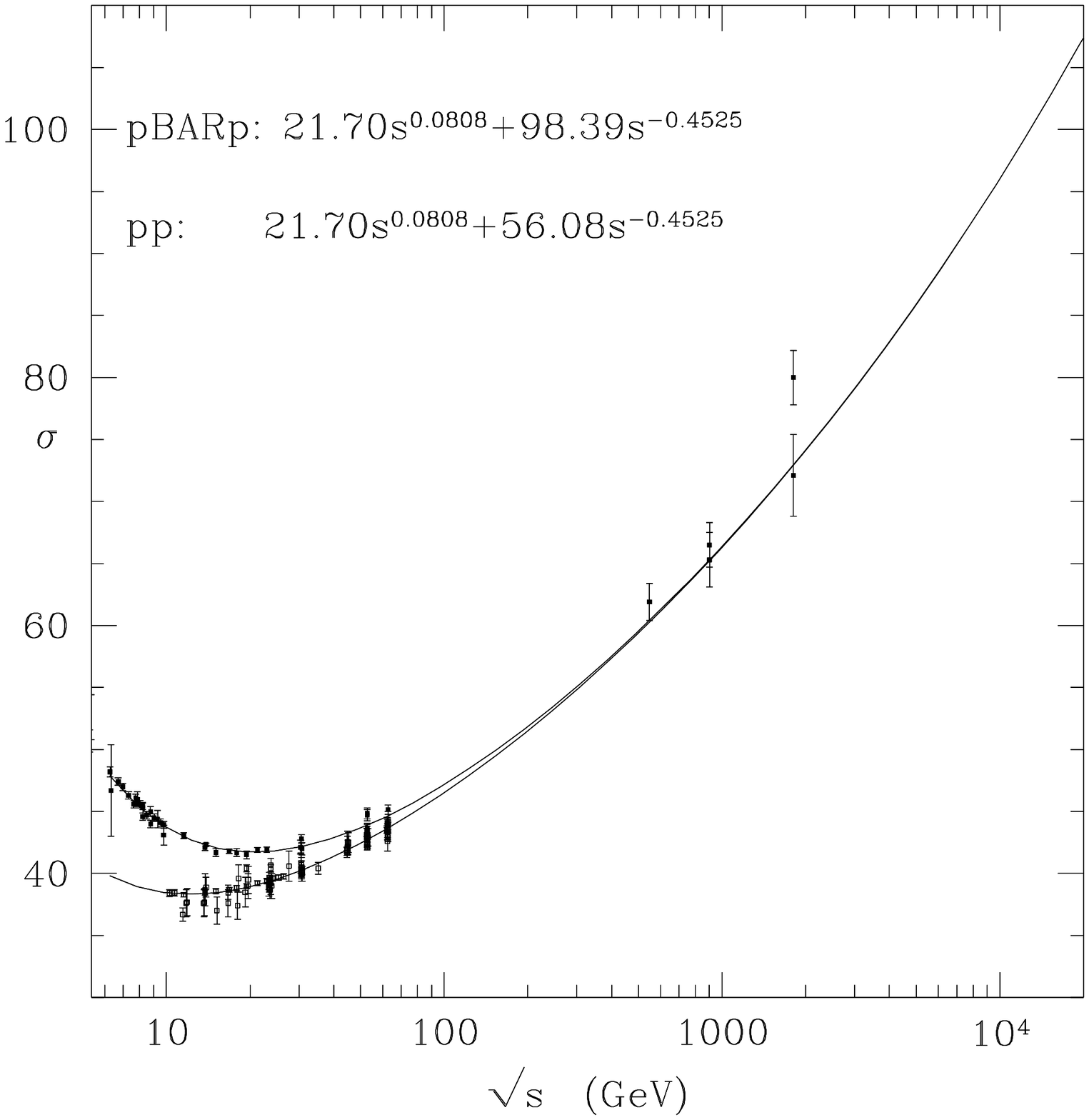}}
\centerline{\rfont Figure 1: $pp$ and
 $\bar pp$ total cross-sections, with simple fits}
\endinsert

Fits\defref\dynamics{
A Donnachie and P V Landshoff, Nuclear Physics B267 (1986)  690
} to elastic scattering data 
suggest that the pomeron trajectory is linear in $t$:
$$
\a (t)=1+\e _0 +\a 't
\eqno(1)
$$
The value $\a '=0.25$ GeV$^2$ was extracted from data more than 20 years
ago\defref\jar{
J A Jaroszkiewicz and P V Landshoff, Physical  Review D10 (1974) 170
} and remains unchanged today. It has a direct effect on the forward
exponential slope $b$ in elastic scattering:
if single-pomeron exchange alone is included then
$$
b=b_0-2\a' |t|\log \a's
\eqno(2)
$$
The extent to which $b$ deviates from linear dependence on $\log \a's$
is therefore a measure of how important are the multiple exchanges.

Note that, even if the exchange of two pomerons is indeed small in
the total cross-section, and therefore also in elastic scattering at
very small $t$, we know that this cannot be so at larger $t$. We cannot
calculate the magnitude of the contribution from this exchange, 
but we do know its general properties.
It is flatter than single-pomeron exchange, and as $s$ increases it steepens
half as quickly. But at $t=0$ it rises twice as fast as single-pomeron exchange.
So, as $s$ increases the point where the two are equal moves to lower and
lower $t$. See figure 2.
One consequence of this is that the shape of the
differential cross-section, as a function of $t$, changes with increasing
energy. It happens that, at Tevatron energy, the two contributions combine
in such a way that a fit $e^{-b|t|}$ with $b$ independent of $t$ is quite
good\defref\esev{
E710 collaboration: N A Amos et al, Physics Letters B247 (1990) 127
}, though this is not true at either lower or higher energies: at
low energies $d\s /dt$ curves downwards in a log/linear plot, while at
very high energies it should curve upwards.

\midinsert
\centerline{{\epsfxsize=50truemm\epsfbox{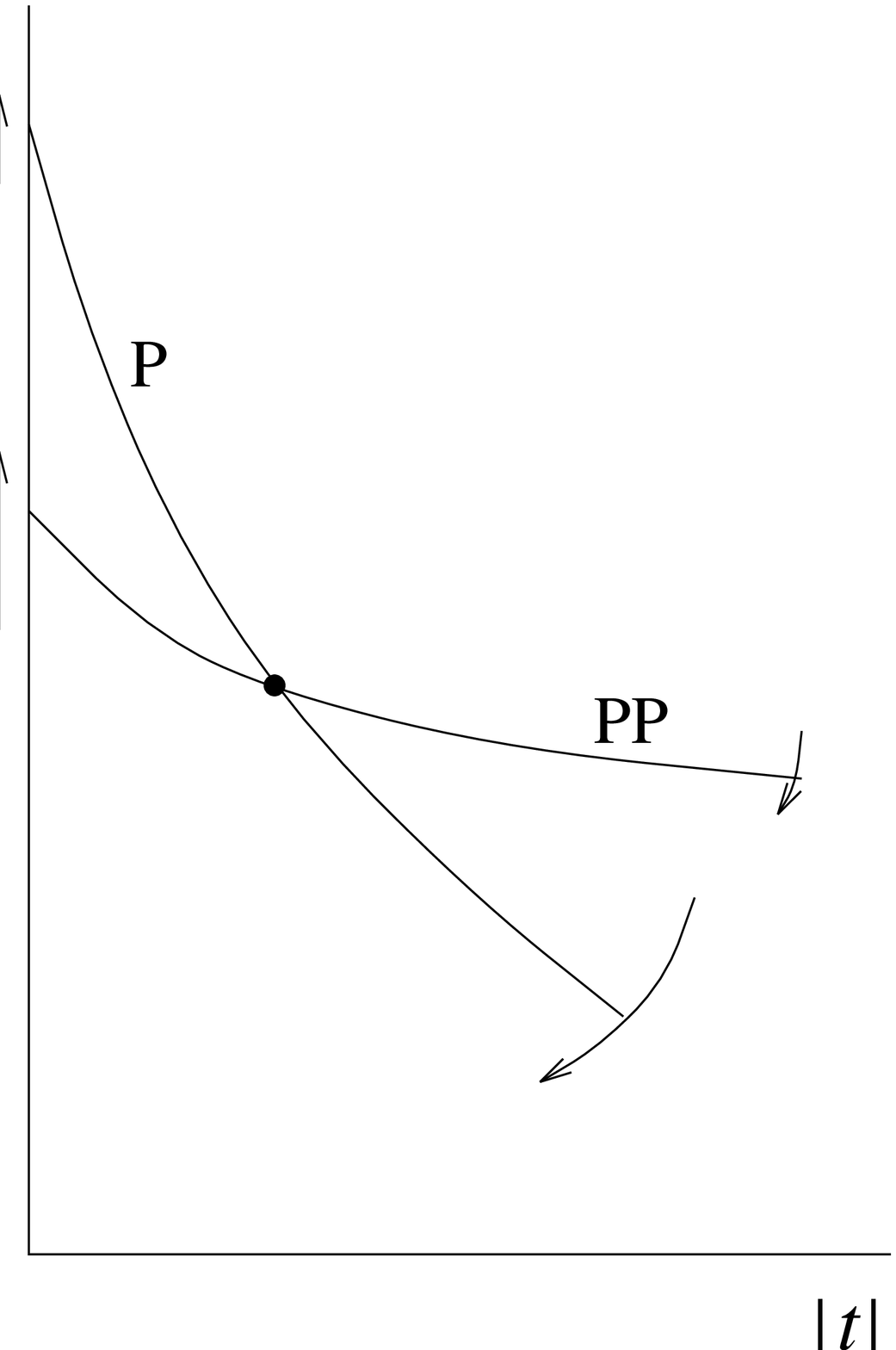}}}\hfill
\vskip -5mm
{\rfont Figure 2: contributions to ${d\s\over dt}$ from single and
double pomeron exchange. The arrows indicate how they change as the energy
increases.}
\endinsert
\bigskip
{\bf Elastic scattering -- medium} {\bfit t}

At ISR energies, the $pp$ elastic scattering differential cross-section
has a dip\defref\dip{
E Nagy et al, Nuclear Physics B150 (1982) 221
}. This dip is around $|t|=1.4$ GeV$^2$ and moves inwards slowly
as the energy increases. It is particularly deep at $\surd s=$ 31 GeV.
In $\bar pp$ scattering the dip is much less pronounced%
\defref\breakstone{
A Breakstone et al, Physical Review Letters 54 (2180) 1985
}, or perhaps even not there at all. The fact that $pp$ and $\bar pp$ 
scattering behave differently in the dip region tells us that there
is a significant $C=-1$ exchange that contributes there. 
Indeed, we have argued\ref{\dynamics} that such an exchange is
important in order to produce a dip at all. This is because the phase
of the single-pomeron-exchange  contribution to the amplitude
at $t$-values near the dip  is very different from 
that of double exchange.
Our belief is\ref{\dynamics} that the imaginary parts of
the single and double exchanges cancel at the dip but then some other 
contribution is needed to cancel the real part of the amplitude, and 
the $C=-1$ exchange is the obvious candidate for this -- it even has
the right sign. It is rather fortuitous that there is a dip at all, since it
needs simultaneous cancellation of the real and imaginary parts, and it
seems that at CERN $\bar pp$ collider energy the ``dip'' has moved so
far inwards that this simultaneous cancellation no longer occurs
and the dip has been replaced by a mere shoulder\defref\ua4{
UA4 collaboration: D Bernard et al, Physics Letters B171 (1986) 142
} 
(it will be interesting to check in $pp$ scattering at RHIC that the
same is true). The LHC data will provide a check on whether single
and double pomeron exchanges are the dominant ones, or whether multiple
exchanges of pomerons are important too. According to figure 2, the
$t$-value at which the magnitudes of the single and double exchanges
are equal continues to move inwards as the energy increases, until 
it reaches a range of values where the phases of the single
and double exchanges are  nearly equal. We estimate%
\ref{\dynamics} that
this will be true at the LHC, and therefore that again there will be a 
noticeable dip -- somewhere near $|t|=0.5$ GeV$^2$.
\bigskip
{\bf Elastic scattering -- large} {\bfit t}

At ISR energies, the $pp$ elastic differential cross section has a strikingly
simple behaviour\defref\data{
W Faissler et al, Physical Review D23 (1981)33
}\ref{\dip} for $|t|$ greater than about 3 GeV$^2$: it is energy-%
independent and behaves at $|t|^{-8}$. This is consistent with simple
three-gluon exchange\defref\larget{
A Donnachie and P V Landshoff, Zeits Physik C2 (1979) 55
}. This same mechanism is likely to be the $C=-1$-exchange mechanism
that helps to give the dip seen at smaller $t$ in the ISR experiments.

It will be interesting to see whether this energy-independence 
at large $t$ survives at the LHC.
It may even be replaced with something more dramatic\defref\drama{
A Donnachie and P V Landshoff, Physics Letters B387 (1996) 637
}. It may be that 3-gluon exchange is replaced with the exchange 
of the exchange of three BFKL pomerons, in which case the large-$t$
differential cross section would {\it rise} rapidly with $\surd s$.
\midinsert
\epsfxsize \hsize\epsfbox{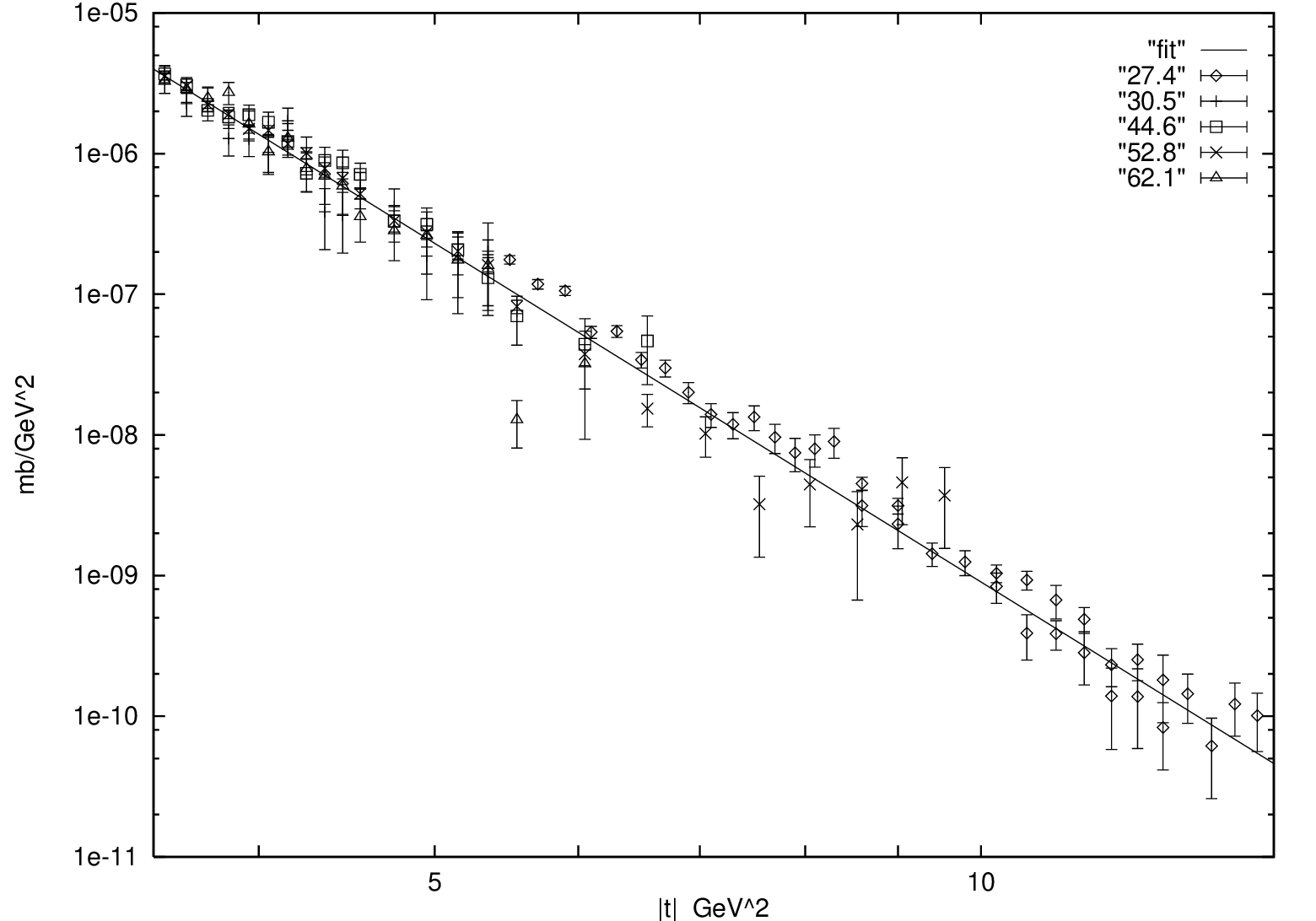}
\centerline{Figure 3: differential cross-section
for $pp$ elastic scattering, with the fit $0.09 |t|^{-8}$}
\endinsert
\bigskip
{\bf Diffraction dissociation}

In diffraction dissociation, an extremely fast proton appears in
the final state, so that necessarily there is a large rapidity gap.
Interest in such events has been revived with the discovery at HERA
that they occur even in deep inelastic electron scattering at high
$Q^2$, which has made it important to try to understand the
puzzling features\defref\goulianos{
K Goulianos, Physics Letters B358 (1995) 379
} of the corresponding $pp$ and $\bar pp$ data.

There is more than a suspicion\defref\schlein{
P Schlein, Proc Workshop on Small-$x$ and Diffractive Physics, Argonne (1996)
} that these puzzling features may not be real,
but rather result from the way the data have been parametrised.
It is important that experimentalists present their complete data,
and not just the results of their own fits to them\defref\cdf{
CDF collaboration: F Abe et al, Physical Review D50 (1994) 5535
}.

For an event to be classified as being diffraction dissociation,
the fractional momentum loss $\xi$ of the fast proton should be less than a
few percent. 
The magnitude of $\xi$ may be calculated from the invariant mass of the
system $X$ of fragments of the other initial-state particle: 
$\xi =M_X^2/s$. Instead of $\xi$, the notation $x_P$ is often used.
If $\xi$ is small enough, the exchanged object should be the
pomeron. If pomeron exchange is described by a simple pole in the complex
$\ell$-plane, it should factorise
\defref\diffdis{
A Donnachie and P V Landshoff, Physics Letters B387 (1996) 637
}:
$$
d^2\s ^{Ap}/dt d\xi=F_{P/p}(\xi ,t)\;\s ^{PA}(M_X^2,t)\quad\hbox{with}\quad
F_{P/p}={9\beta _0^2[F_1(t)]^2\over 4\pi}\xi ^{1-2\a (t)}
\eqno(3)
$$
where $\beta _0$ is the coupling strength of the pomeron to a quark
and $F_1(t)$ is the proton's elastic form factor.

One issue in pomeron physics is whether the pomeron trajectory has on it
particles, as do other Regge trajectories. If so, the popular belief is
that these particles should be glueballs, and indeed there is\defref\wa91{
WA91 collaboration: S Abatzis et al, Physics Letters B324 (1994) 509
}
a 2$^{++}$
candidate for the first of these, at a mass a little less than 2 GeV.
But even if there are particles on the pomeron trajectory,
when it is exchanged near $t=0$ the pomeron cannot be said
to be a particle. Nevertheless, the factorisation (3) -- if it is valid --
makes pomeron
exchange very similar to particle exchange: the factor $\s ^{PA}(M_X^2,t)$
may be thought of as the cross-section for pomeron-$A$ scattering. When
its subenergy $M_X$ is large, it should have much the same power behaviour as
the hadron-hadron total cross-sections: a combination of
$(M_X^2)^{0.08}$ and $(M_X^2)^{-0.45}$ with $t$-dependent coefficients.
The relative mix of these two contributions must be determined by experiment
and there are not yet enough good data for this have been done definitively.
In triple-Regge-exchange language, this amounts to saying that the
relative strengths of the PPP and PPR terms is not well known.

But there are other complications: the exchange may not be the pomeron.
Simple pomeron exchange may be contaminated in two ways. If $\xi$ is not
small enough, one must add in a contribution from $\rho,\o ,f, a$ exchange,
or even $\pi$ exchange when $t$ is close to 0. 
Parametrisations\defref\robroy{
D P Roy and R G Roberts, Nuclear Physics B77 (1974) 240
}\ref{\diffdis} of ISR, CERN collider and HERA\defref\hh1{
H1 collaboration: C Adloff et al, hep-ex/9702003
} 
data suggest that contributions from nonleading exchanges
are large. In triple-Regge language, these are the PRP, PRR, RRP and RRR
terms. It is evident that a great deal of data will be needed to fix them
all well.

If one integrates (3) down 
to some fixed $M_X^2$, the resulting cross-section for diffraction
dissociation behaves as $s^{2\e _0}$, and so unless something else intervenes
it would become larger than the total cross-section\defref\ss{
G A Schuler and T Sjostrand, Physics Letters B300 (1993) 169\h
E Gotsman, E M Levin and U Maor, Physical Review D49 (1994) 4321
}\ref{\goulianos}.
As $s$ increases at fixed $M_X^2$, one is probing larger and larger values
of $1/\xi$, so one expects that the same happens as in the total cross-section:
the exchange of two pomerons becomes important and moderates the rising
contribution from single exchange. But the {\sl simplest assumption} is that
this matters only at very small $\xi$.

Notice that the theory leads us to expect that 
adding all these exchanges  should give us all the nonleading powers of
$1/\xi$: there should be no other appreciable ``background''.
Note also that adding together all the  exchanges will break factorisation.
\bigskip

{\bf Semisoft processes}

Despite the success of this simple picture of the pomeron in describing a
wide range of soft hadronic data, problems do arise when it is applied to
processes in which there is a hard scale, even if that scale is still
rather soft. The most obvious is deep inelastic scattering. At small $x$ 
and small $Q^2$ the structure function $\nu W_2$ is governed by Regge 
theory\defref\lps{
P V Landshoff, J C Polkinghorne and R D Short, Nuclear Physics B28 (1970) 210
} giving a sum of terms $(1/x)^{\alpha(0)-1}$. The NMC data\defref\nnmc{
NMC collaboration: P Amaudruz et al, Physics Letters B295 (1992) 159
}
at moderate $Q^2$ and not-too-small $x$ show that $\nu W_2$ contains such 
Regge terms: a slowly varying term from soft pomeron exchange and close to
$\surd s$ from $f$, $a$ exchange. A simple fit\defref\simple{
A Donnachie and P V Landshoff Z Physik C61 (1994) 139
} to the small-$x$ NMC 
data analogous to the fits\ref{\total} to the total hadronic cross sections
gives an excellent description of the data and the predictions of the fit
are in remarkably good agreement with the subsequent E665 data\defref\e665{
E665 collaboration: M R Adams et al, Physical Review D54 (1996) 3006
}  over
the same $Q^2$ range but at smaller values of $x$ than used in the original
fit. However at yet smaller $x$ comparison with the HERA data\defref\hera{
H1 collaboration: S Aid et al, Nuclear Physics B470 (1996) 3\h
ZEUS collaboration: M Derrick et al, Z Physik C72 (1996) 399
}
fails for values of $Q^2$ not much above 0.5 GeV$^2$. The implications are 
either that perturbative QCD is applicable at this surprisingly small value
of $Q^2$ or that the non-perturbative processes are more complex than the
simple picture allows.

The latter view is one adopted, for example, by Maor\defref\maor{
U Maor, Proc  Workshop on Quantum Infrared Physics, Paris, 1994, ed 
H M Fried and B Muller, World Scientific, p 321
}  in the context
of the eikonal model. The principal effect is to make significant changes
in the relative energy dependence of the total, elastic and diffractive 
dissociation cross sections as the energy scale at which screening
effects become appreciable differs in each case. It is thus possible to 
fit simultaneously the total cross section and the CDF diffraction
dissociation data\ref{\cdf}. Further, it is possible to take a large value
for $\epsilon$ which is reduced to an effective low value in hadron
hadron and in photon hadron reactions by the strong unitarity corrections. 
However for virtual photons these corrections start to diminish at some
moderate value of $Q^2$ and so can incorporate the transition seen in deep
inelastic scattering\defref\abram{
H Abramowicz, E M Levin, A Levy and U Maor, Physics Letters B269 (1991) 465\h
A Capella, A Kaidalov, C Merino and J Tran Thanh Van, 
Physics Letters B337 (1994) 358}.

\bigskip\bigskip

{\sl This work was supported by the
EC Programme ``Training and Mobility of Researchers", Network
``Hadronic Physics with High Energy Electromagnetic Probes",
contract ERB FMRX-CT96-0008, and by PPARC}
\vfill\eject
\medskip\immediate\closeout\rfile\writestoppt
\parskip=0pt
{{\bf References}}\vskip 1mm{\frenchspacing%
\parindent=20pt\escapechar=` \input refs.tmp\bigskip}\nonfrenchspacing
\bye